%% file: AoI_GlobeCom.tex
\documentclass[10pt,conference]{IEEEtran}
\usepackage{setspace}
\usepackage{amsmath}
\usepackage{acronym}
\usepackage{bbm}
\usepackage[dvips]{color}
\usepackage{comment}
\usepackage{todonotes}
\usepackage{epsf}
\usepackage{siunitx}
\usepackage{epsfig}
\usepackage{times}
\usepackage{epsfig}
\usepackage{graphicx}
\usepackage{bbold}
\usepackage{geometry}
\usepackage{mathrsfs}
\usepackage{amssymb}
\usepackage{pdfpages}
\usepackage{epstopdf}
\usepackage{tcolorbox,bbold}
\usepackage{algorithm,algorithmicx,algpseudocode}
\makeatletter
\newcommand{\algmargin}{\the\ALG@thistlm}
\makeatother
\algnewcommand{\parState}[1]{\State%
	\parbox[t]{\dimexpr\linewidth-\algmargin}{\strut #1\strut}}

\usepackage[nomain,acronym,shortcuts]{glossaries}
\newcommand*{\acro}[3][]{\newacronym[#1]{#2}{#2}{#3}}
\include{acronyms}

\usepackage{amssymb}
\usepackage{url}
\usepackage{dsfont}
\usepackage{lettrine} 
\usepackage{amsmath,epsfig,amssymb,algorithm,algpseudocode,amsthm,cite,url}
\IEEEoverridecommandlockouts
\usepackage{subcaption}
\allowdisplaybreaks
\usepackage{csquotes}

\usepackage{verbatim}
\usepackage[english]{babel}
\usepackage{amsmath,amssymb}

\usepackage[justification=centering]{caption}
\usepackage{verbatim}

\newtheorem{theorem}{\bf Theorem}

\IEEEoverridecommandlockouts
\begin{document}
	\title{On the Ruin of Age of Information in Augmented Reality over Wireless Terahertz (THz) Networks \vspace{-0.35cm}}
	\author{\IEEEauthorblockN{Christina Chaccour and
			Walid Saad\\}
		\IEEEauthorblockA{Wireless@VT, Bradley Department of Electrical and Computer Engineering, Virginia Tech, Blacksburg, VA USA,}
		\IEEEauthorblockN{Emails:\{christinac, walids\}@vt.edu.}
		\thanks{This research was supported by the Office of Naval Research (ONR) under MURI Grant Grant N00014-19-1-2621 and by the National Science Foundation under Grant CNS-1836802.}
	\vspace{-1cm}}
	\maketitle
	%
	

	\begin{abstract}
		Guaranteeing fresh and reliable information for \ac{AR} services is a key challenge to enable a real-time experience and sustain a high \ac{QoPE} for the users. In this paper, a \ac{THz} cellular network is used to exchange rate-hungry \ac{AR} content. For this network, guaranteeing an instantaneous low \ac{PAoI} is necessary to overcome the uncertainty stemming from the \ac{THz} channel. In particular, a novel economic concept, namely, the risk of ruin is proposed to examine the probability of occurrence of rare, but extremely high \ac{PAoI} that can jeopardize the operation of the AR service. To assess the severity of these hazards, the \ac{CDF} of the \ac{PAoI} is derived for two different scheduling policies. This \ac{CDF} is then used to find the probability of maximum severity of ruin \ac{PAoI}. Furthermore, to provide long term insights about the \ac{AR} content's age, the average \ac{PAoI} of the overall system is also derived. Simulation results show that an increase in the number of users will positively impact the \ac{PAoI} in both the expected and worst-case scenarios. Meanwhile, an increase in the bandwidth reduces the average \ac{PAoI} but leads to a decline in the severity of ruin performance. The results also show that a system with preemptive \ac{LCFS} queues of limited size buffers have a  better ruin performance ($\SI{12}{\%}$ increase in the probability of guaranteeing a less severe \ac{PAoI} while increasing the number of users), whereas \ac{FCFS} queues of limited buffers lead to a better average \ac{PAoI} performance ($\SI{45}{\%}$ lower \ac{PAoI} as we increase the bandwidth).
	\end{abstract}
	\vspace{0cm}
	\vspace{-.35cm}
\section{Introduction}
\begin{figure}
	\vspace{-.25cm}
	\centering
\includegraphics[scale=0.24]{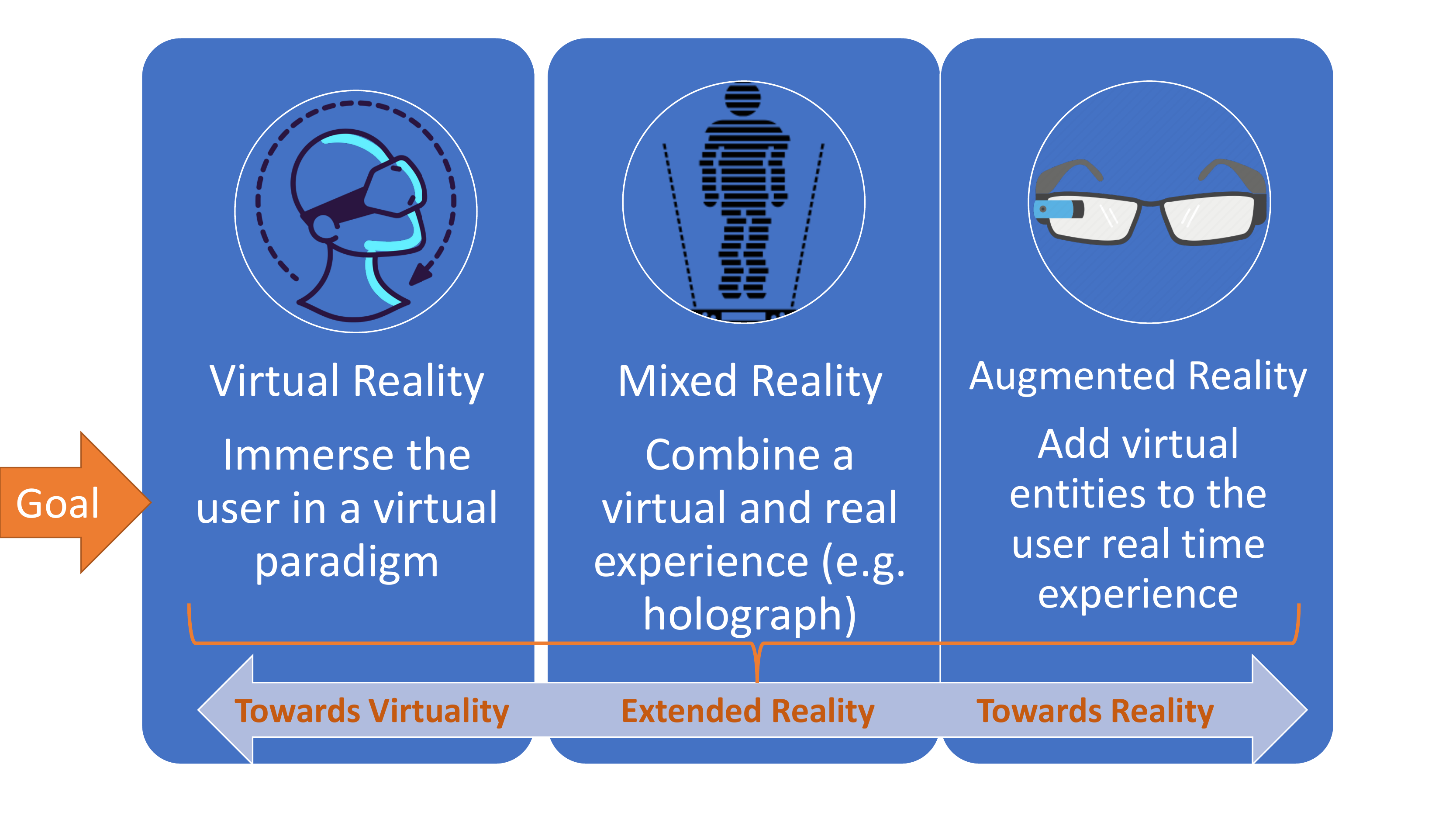}
\vspace{-0.3cm}
\caption{\small{Illustrative example of the \ac{XR} spectrum}}
\label{fig::XR}
\vspace{-0.6cm}
\end{figure}

The emergence of wireless \ac{XR} will yield a radical paradigm shift from conventional network designs aiming to fulfill a \ac{QoS}, towards ones targeting a new concept dubbed \emph{\acrfull{QoPE}} \cite{saad2019vision}. To maintain a high \ac{QoPE}, the wireless network should be capable of soliciting its users' five senses. \ac{XR} encompasses \acrfull{AR}, \ac{VR}, and \ac{MR} as shown in Fig.~\ref{fig::XR}, whose applications will offer a wide range of experiences with different goals. On the virtual end of the spectrum, \ac{VR}'s main goal is to fully immerse the user in a virtual experience, whereas \ac{AR} aims to supplement reality with virtual objects.
Henceforth, to ensure \emph{immersion}, \ac{XR} services necessitate  visual and haptic requirements that are translated onto a wireless downlink at \ac{HRLLC} as investigated in \cite{chaccour2020can}. Clearly, to exchange \ac{XR} content at high rate on the downlink and uplink, it is natural to adopt frequency bands beyond \ac{mmWave}, making the communication at \acrfull{THz} band necessary.\\
\indent Nevertheless, an \ac{HRLLC} downlink in \ac{AR} would be serving outdated requests if it is not supplemented with a call for \emph{fresh} information at the uplink. 
%
The freshness of this uplink information determines the timeliness of \ac{AR} requests to the server that generates the \ac{AR} experiences. 
Since \ac{AR} will be used  in critical applications (e.g. Google Glass is expected to be heavily deployed by numerous industries ranging from manufacturing all the way to surgical operations \cite{john2020review})  a single disruption caused by an \emph{extreme event}, such as highly outdated/delayed uplink data or an unusual traffic pattern, can lead to substantial hazards. For instance, an obsolete \ac{AR} request in a biological lab translates into a biohazard that poses threats to the lives of individuals or the security of property. Thus, apart from disrupting the user's \ac{QoPE}, outdated information in an \ac{AR} network will lead to tremendous faults and great risks. To address this challenge, the freshness of information can be quantified by the concept of \ac{AoI}. For wireless \ac{AR} users, \ac{AoI} depends on both the generation and transmission of \ac{AR} content while capturing the receiver information freshness. Moreover, given that \ac{THz} frequencies are necessary to transmit rate-hungry \ac{AR} content, it is imperative to understand whether an \ac{AR} wireless network over \ac{THz} can sustain an instantaneous low \ac{AoI} to provide the user the promised \ac{QoPE}.
\vspace{-.27cm}
\subsection{Prior Works}
\vspace{-.15cm}
The concept of \ac{AoI} has recently seen a surge in literature such as in \cite{sun2017update, zhou2019minimum, yates2018age}. In \cite{sun2017update}, the optimal control of information updates from a source node to a destination was studied. The work in \cite{zhou2019minimum} studied the problem of minimizing non-uniform status packet sizes for IoT monitoring systems. The authors in \cite{yates2018age} examined the \ac{AoI} timeliness metric through a variety of queuing systems. Meanwhile, the concept of  \acrfull{PAoI} has been studied in \cite{costa2016age, chen2016age, huang2015optimizing}. In \cite{costa2016age}, the authors characterized the average age and the \ac{PAoI} for different scheduling policies.
 The authors in \cite{chen2016age} investigated the expected \ac{PAoI} expressions under different scheduling policies. 
 The work in \cite{huang2015optimizing} studies the tradeoff between frequency of status updates and queuing delay.
 However, these works \cite{sun2017update, zhou2019minimum, yates2018age, costa2016age, huang2015optimizing, chen2016age} do not examine the behavior of \ac{AoI} in providing fresh information to critical and stringent \ac{XR} applications. In fact, the works in \cite{sun2017update, zhou2019minimum, yates2018age, costa2016age, huang2015optimizing, chen2016age} do not examine the \emph{instantaneous} aging process necessary to deliver a real-time \ac{AR} experience, thus, they do not shed light on extreme events with a high \ac{AoI}. Such events lead to substantial risks in critical \ac{AR} applications and require a study of \emph{the risk of ruin}, i.e., the likelihood of hazardous damages caused by exceeding the \ac{PAoI} measure. Such an analysis is fundamentally important for a better understanding of the uncertain \ac{THz} channel.\\
\indent The main contribution of this paper is a ruin-aware novel performance analysis in terms of achievable average peak \ac{AoI}, worst-case \ac{AoI}, and reliability for a cellular network operating at \ac{THz} frequencies and servicing \ac{AR} users. The ultimate goal is to assess how and when a \ac{THz} network can deliver minimal peak \ac{AoI}, in terms of the \ac{THz} data rate, depending on the adopted network queuing policies. In particular, we introduce a model for a wireless \ac{AR} service that is deployed using \ac{THz} operated \acp{RIS}. In the studied model, each \ac{AR} user sends a request to an \ac{RIS} with \ac{MEC} capabilities, to solicit new \ac{AR} content. Given the stringent \ac{QoPE} requirements of \ac{AR} services and the uncertainty of the \ac{THz} channel, it is fundamental to examine the expected and the ``ruin'' \ac{PAoI}, which captures the worst-case information obsoleteness. Subsequently, to guarantee that the \ac{PAoI} does not violate the \ac{QoPE} we derive the \acrfull{CDF} of the maximum severity of ruin, i.e., the distribution characterizing the severity of \ac{PAoI} exceedances. Moreover, we derive the average \ac{PAoI} to provide long term insights about the \ac{AR} response. To our best knowledge, \emph{this is the first work that analyzes the ruin oriented \ac{PAoI} achieved by \ac{AR} services over a \ac{THz} cellular network.} Simulation results show that an increase in the number of users positively impacts the \ac{PAoI} both in the expected and worst-case, while an increase in the bandwidth reduces the average \ac{PAoI} but leads to a decline in the severity of ruin performance.  \\
\indent The rest of this paper is organized as follows. The system model and problem formulation are presented in Section II. The ruin-oriented peak \ac{AoI} analysis is performed in Section III. Section IV presents the simulation results. Finally conclusions are drawn in Section V.
\vspace{-.25cm}
\section{System Model and Problem Formulation}\label{1-1} 
Consider the uplink\footnote{\noindent \vspace{-.1cm}The downlink of AR transmission is assumed to follow an arbitrary \ac{THz} scheme and is outside of the scope of this chapter} of an \ac{RIS}-based wireless network in a confined indoor area, servicing a set $\mathcal{U}$ of $U$ mobile wireless \ac{AR} users via a  set $\mathcal{B}$ of $B$ \acp{RIS} acting as \ac{THz} operated \acp{BS}. In particular, \acp{RIS} are a scaled-up version of massive \ac{MIMO} that supplement existing walls and structures with wireless capabilities. Thus, \acp{RIS} increase the likelihood of a near-field communications through a \ac{LoS} path that is fundamental for reliable \ac{THz} communications\footnote{\noindent The analysis of the probability of blockage is outside the scope of this work, a guaranteed \ac{LoS} link is assumed since we use \acp{RIS} \cite{basar2019wireless} deployed on existing walls and structures at a close proximity of the users.}. Moreover, the \ac{AR} users are mobile and may change their locations and orientations at any point in time. We consider continuous time slots indexed by $t$ with fixed duration $\tau$. Each \ac{RIS} is a \ac{BS}, that is provided with a feeder (antenna) with a corresponding transmit power denoted by $p$. Hence,  the transmitted data is encoded onto the phases of the signals reflected from different reconfigurable meta-surfaces that compose the \ac{RIS} \cite{basar2019wireless}. Henceforth, if the \ac{RIS} consists of $N$ meta-surfaces whose reflection phase can be optimized independently, then an $N$-stream virtual \ac{MIMO} system can be realized by using a single \ac{RF} active chain \cite{basar2019wireless}. We assume that the \ac{RF} source is close enough to the \ac{RIS} surface so that the transmission between each pair of \ac{RF} source and \ac{RIS} is not affected by fading \cite{basar2019wireless}.
\vspace{-0.25cm}
\subsection{Wireless Capacity}
We consider an arbitrary \ac{AR} user $u$ in $\mathcal{U}$ that is at a constant distance $d_{ub}$ from its respective \ac{RIS} $b$. Moreover, as a byproduct of directional beamforming, propagation differences, and \ac{RIS} deployment, the considered network will be noise limited. Furthermore, deploying \acp{RIS} enables the desired channel to become a \ac{LoS} channel \cite{jung2019reliability}. Subsequently, to guarantee a \ac{LoS} link for every \ac{AR} user, we assume that the \ac{RIS}-\ac{AR} user association is performed by a centralized controller\cite{chaccour2020risk}. In particular, the electromagnetic response of the $N$ meta-surface elements is programmed by generating input signals that tune varactors and change the phase of the reflected signal~\cite{huang2019reconfigurable}. Let $\Theta_{ub,t}=[\theta_{ubn,t}]_{N \times 1}$ be the phase shift vector of \ac{RIS} $b$ , with respect to \ac{UE} $u$, at time slot $t$, where $\theta_{ubn,t} \in \Theta$, $n$ is the index corresponding to the meta-surface of each \ac{RIS}, and $\Theta=\{-\pi+\frac{2k\pi}{K-1}|k=0,1,...,K-1\}$. $K$ is the number of possible phase shifts per meta-surface element. Thus, the random channel gain between \ac{AR} \ac{UE} $u$ at time slot $t$ is given by \cite{chaccour2020can}: $h_{ub,t}=\big(\frac{\lambda}{4\pi d_{ub,t}}\big)^2\big(e^{-k(f)   d_{ub,t}}\big)^2,$ where  $k(f)$ is the overall molecular absorption coefficient of the medium at the \ac{THz} band, and $f$ is the operating frequency. Let $\psi_{ubn,t}$ be the phase shift vector between \ac{AR} \ac{UE} $u$ and the metasurface $n$ of \ac{RIS} $b$ at time $t$. Since the coherence bandwidth at \ac{THz} is inherently large due to the delay spread and temporal broadening effects as shown in  \cite{kokkoniemi2014frequency} and \cite{hantest2017ma}, we can assume that the rate shows an invariant behavior across the \ac{THz} bandwidth for all the distances considered. Hence, for a given reflection phase shift vector, $\theta_{ub,t}$, the transmission rate from \ac{AR} \ac{UE} $u$ to \ac{RIS} $b$ will be (under a an approximate average \ac{SINR} value across the \ac{THz} band):
\vspace{-.25cm}%
\begin{equation*}
 {R}_{ub,t}= W \log_2 \left( 1+\frac{p   {h}_{ub,t} |\sum_{n=1}^{N} e^{(\theta_{ubn,t}-\psi_{ubn,t})j}|^2}{N({  d_{ub,t}},p,f)} \right) ,
\label{Received bitrate}
\end{equation*}
where $N({  d_{ub,t}},p,f)=N_0+\sum_{b=1}^{B}pA_0d_{ub,t}^{-2}(1-e^{-K(f)  d_{ub,t}})$,  $N_0=\frac{W \lambda^2}{4 \pi} k_B T_0$, $k_B$ is the Boltzmann constant, $T_0$ is the temperature in Kelvin,  $A_0=\frac{c^2}{16 {\pi}^2 f^2}$, and $c$ is the speed of light \cite{zhang2018analytical, chaccour2020can}.
Note that the optimal choice for $\theta_{ubn,t}$ for every \ac{RIS} association is equal to $\psi_{ubn,t}$, thus maximizing the rate ${r}_{ub,t}$, as shown in \cite{basar2019wireless}.\\
\indent Furthermore, the rate of \ac{AR} content transmission over the \ac{THz} link between \ac{AR} user $u$ and \ac{RIS} $b$, assuming a constant image size $M$, is given by: $r_{ub,t}=\frac{R_{ub,t}}{M}$. 
\vspace{-.2cm}
\subsection{Age of Information}
\vspace{-.1cm}
The \ac{AoI} is a key performance metric to quantify the freshness of the status information update at the receiver defined as the time elapsed since the most recent \ac{AR} content update delivered at the \ac{MEC} server. By definition, the \ac{AoI} at the receiver at the beginning of time slot $t$ is given by: $A_{ub,t}=t-U_t^B$, where $U_t^B$ is the time stamp of the freshest \ac{AR} content update that was delivered to the \ac{RIS} before $t$. Moreover, while the use of average \ac{AoI} as a freshness metric might guarantee a good performance on the long run, for real-time applications such as AR, it is necessary to examine the \emph{instantaneous} aging of information, so as to, capture real-time disruptions to the \ac{AR} experience. Subsequently, our interest lies in evaluating the distribution of the \ac{PAoI} and its worst-case performance. In \cite{kaul2012status}, the authors have shown that the \ac{AoI} is minimized when the queuing follows a \acrfull{LCFS} scheme in contrast to an \ac{FCFS} scheme, also, an \ac{LCFS} with preemption performed better than an \ac{LCFS} scheduling without preemption. Nevertheless, this hypothesis is true for queuing schemes with one source, or when all the transmitters are sending out status updates about the same situation. In other words, the freshest update among all sources is always the most desirable update to the receiver.
Also, this analysis was solely based on average \ac{AoI} values and overlooked the instantaneous performance.\\
\indent In our model, while one user is immersed into one \ac{AR} session and needs to convey its freshest information to the \ac{MEC} server, another \ac{AR} user might be undergoing the same process. On the one hand, we need to ensure a fair scheduling among users and deliver every request freshly to the server, thus, doing so with an \ac{FCFS} queue allows this process. On the other hand, in case the user needs to urgently deliver a new request, it would be best for the server to ignore the previous requests coming from that user and respond to the most recent one. For example, in a manufacturing plant, if an \ac{AR} user commits a fault and needs an urgent \ac{AR} content assistance in solving the situation, an \ac{LCFS} queuing policy would be the most appropriate for this situation. Otherwise, not only outdated information will be sent, but \emph{risk-inducing outdated information} will be sent to the \ac{AR} user.
	\begin{figure} [t]
				\vspace{-0.4cm}
				\centering
		\includegraphics[scale=0.43]{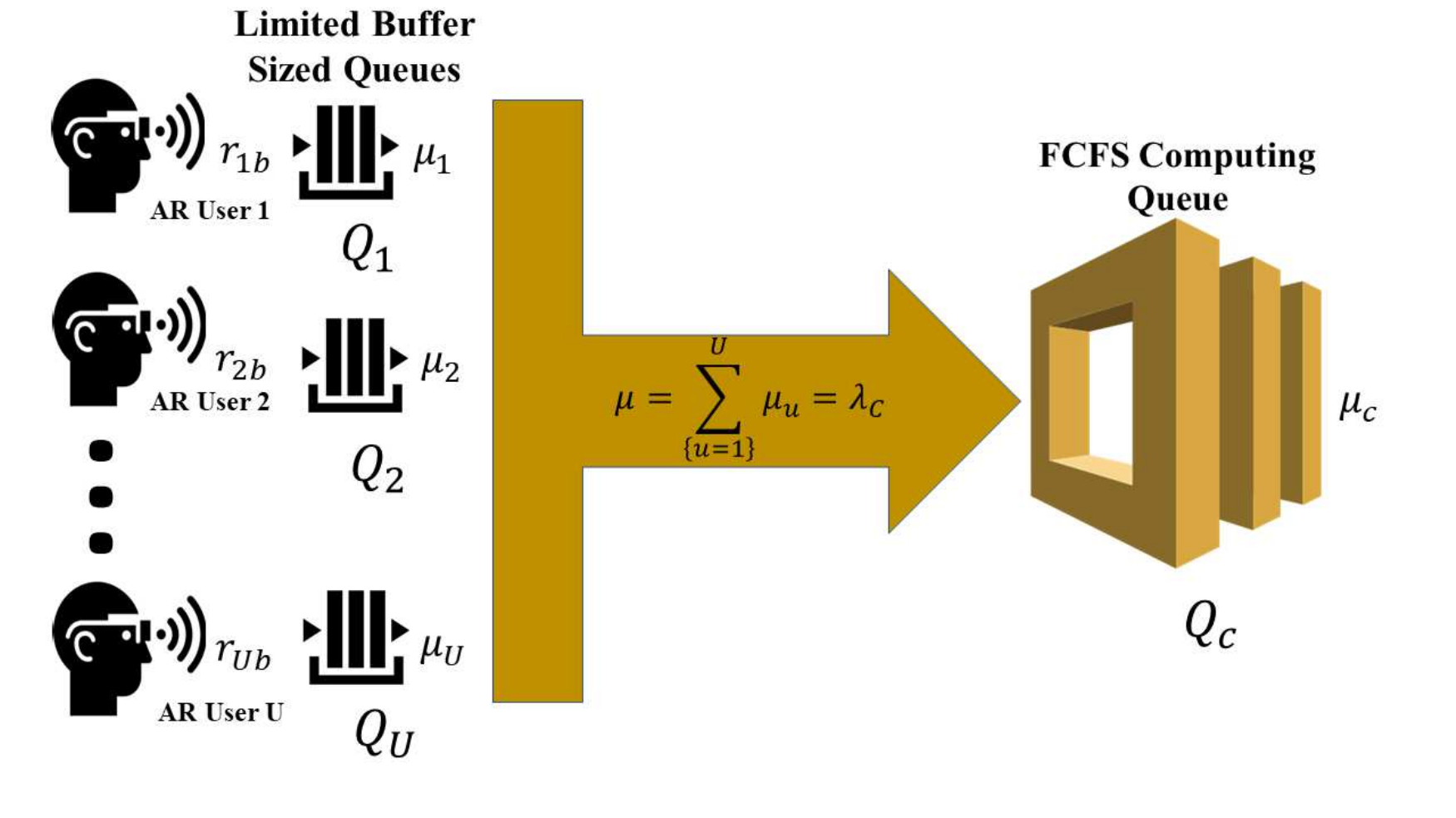}
		\vspace{-.50cm}
		\caption{\small{Illustrative example of our queuing model}}
		\label{fig:sys_model}
		\vspace{-0.6cm}
\end{figure}
Therefore, to scrutinize this problem, we propose to split the queuing system into two as shown in Fig.~\ref{fig:sys_model}. Splitting the queuing system allows us to diagnose the complex scheduling situation and to characterize the freshness of the information as a function of fair and priority scheduling policies. Here, similar to \cite{yates2018age} and \cite{javani2019age}, we assume that the \ac{AR} update arrivals are modeled as a \ac{i.i.d.} \ac{PPP} with temporal mean rate $r_{ub}=\mathbb{E}\left[ r_{ub,t}\right]$. These \ac{AR} updates are processed and halted in limited sized M/M/1/x  queues (x is the buffer size) and then sent to a common \ac{FCFS} queue, where a new \ac{AR} content is generated as a response to the status update. Such limited sized queues that provide halting allow us to inject a sense of prioritization to users without leading to an unfair scheduling policy as in with a complete \ac{LCFS} queue, i.e., priority only when necessary.\\
\indent According to Burke's Theorem \cite{queuingtheory}, when the service rate is larger than the arrival rate for an M/M/1 queue, then the departure process at steady state is a Poisson process with the same arrival rate. Hence, the arrival of requests to $Q_C$ also follows a Poisson process with rate $\lambda_C =\sum_{u=1}^U\mu_u$, where $\mu_U$ is the service rate of each limited buffer sized queue. We also assume that the computing follows a Markovian process and thus, $Q_c$ is an M/M/1 queue with service rate $\mu_C$. Next, we will analyze two types of limited sized queues with different dynamics and buffers. Given the risks pertaining to outdated \ac{AR} requests, we use a financial concept, the \emph{risk of ruin}, defined as an insurer's likelihood of losing an investment capital \cite{dickson2016insurance} and apply it to critical \ac{AR} requests to characterize the likelihood of hazardous damages based on the aging process. In particular, we derive the ruin of going above a \ac{PAoI} threshold set for this network and the \ac{CDF} of maximum severity of ruin. We then compare this ruin-oriented result performance to the average \ac{PAoI} performance.
\vspace{-.30cm}
\section{Ruin-Oriented Peak AoI Analysis}
\vspace{-.05cm}
In this section, we examine the instantaneous \ac{PAoI} and use it to derive the probability of ruin of \ac{PAoI}. Furthermore, we evaluate the average \ac{PAoI} pertaining to the overall queuing system and compare its performance to the worst-case \ac{PAoI}, hence shedding light on the complimentarity of their insights and necessity of both average and tail analyses.
\vspace{-.20cm}
\subsection{Instantaneous and Ruin AoI} 
\vspace{-.15cm}
Here, we analyze the performance of the peak \ac{AoI} of \ac{AR} generated content at \ac{THz} frequency with respect to a limited size \ac{FCFS} M/M/1/2 queue and an \ac{LCFS} M/M/1/2* queue, which allows preemption only in waiting, i.e., once an \ac{AR} content is sent to the computing queue it cannot be preempted. These two types will be evaluated based on the ruin of peak \ac{AoI} and its maximum severity.\\
\indent Hereinafter, the term \emph{reliability} is defined as the ability of the \ac{THz} network to maintain low \ac{PAoI}. Moreover, analyzing the reliability of the network in terms of \ac{PAoI} needs to be performed \emph{instantaneously} and needs to account for the extreme events that will violate the \ac{QoPE} target performance. In the light of this, we propose a novel notion of ruin-based reliability analysis developed using notions from the powerful actuarial sciences and economics framework \cite{dickson2016insurance, michaud1996estimating, liu2016theory}. From economics, the probability of ruin in finite time is defined as:\vspace{-.15cm}
\begin{equation*}
\vspace{-.2cm}
\Psi(u,t)=\operatorname{Pr}(U(s)<0 \quad \text { for some } s, 0 < s \leq t),
\end{equation*}
where $U(s)=u+c t-S(s)$, $u$ is the insurer's surplus at time 0 and $c$ is the insurer's rate of premium income per unit time, which we assume to be received continuously. Thus, $\Psi(u,t)$ is the probability that the insurer's surplus falls below zero at some time in the future, i.e., the claims outgo exceeds the initial surplus plus premium income. This is a probability of ruin in continuous time, we can similarly define a discrete time ultimate ruin probability. Moreover, the insurer's surplus in the financial domain, translates onto  reliability in a wireless network. Moreover, a duality exists between the queuing dynamics of an M/G/1 queue and the risk of ruin as shown in \cite{michaud1996estimating}. Elaborating on this duality, we can define the system reliability in terms of \ac{PAoI} and let $\Psi(u,t)$ characterize its \ac{CDF}. Subsequently, we will find this distribution and use it to derive the maximum severity of ruin for each one of the scheduling schemes considered.\\
\indent We first evaluate the \ac{PDF} of the peak \ac{AoI} for both scheduling configurations. To do so, we define $P(\nu)=\frac{\mu_u}{r_{ub}+\mu_u}$ as the steady state probability as the system is empty, this probability is used to compute the peak \ac{AoI} \ac{PDF}, whose conditional is given by: $\Phi(a | \nu)=\Phi(\tau| \nu) * \Phi(y | \nu)$, where $a$ denotes the value of peak \ac{AoI} , $\tau$ is the time spent in the system, and $y$ is the interdeparture time. Thus, knowing the conditional \acp{PDF} from \cite{costa2016age}, we can write the marginal \ac{PDF} by using Bayes' theorem as follows: 
\vspace{-0.25cm}
\begin{align}
\label{fcfs}
&\Phi_F(a)=\Phi_F(a|\nu)P(\nu)+\Phi_F(a|\bar{\nu})P(\bar{\nu}), \\
\begin{split}
\vspace{-.1cm}
&=\left(\frac{\mu_u}{r_{ub}-\mu_u}\right)^{2}r_{ub}\left(e^{-r_{ub} a}- e^{-\mu_u a}+(r_{ub}-\mu_u) \times \right. \\
&\quad \left. a e^{-\mu_u a}\right) \left( \frac{\mu_u}{r_{ub}+\mu_u}\right) + \frac{1}{2} a^{2} \mu_u^{3} e^{-\mu_u a}\left( \frac{r_{ub}}{r_{ub}+\mu_u} \right).\nonumber 
\vspace{-1.5cm}
\end{split}
\end{align}
 Following a similar procedure for the M/M/1/2* queue, the peak \ac{AoI} \ac{PDF} of each queue is given by:
 \vspace{-.1cm}
 \begin{align}
 \label{preemption}
  &\Phi_L(a)=\Phi_L(a|\nu)P(\nu)+\Phi_L(a|\bar{\nu})P(\bar{\nu}), \nonumber \\
 \end{align}
 \vspace{-1.1cm}
 \footnotesize {
 \begin{align}
 \begin{split}
 \vspace{-.25cm}
& =\left( \left(r_{ub}(r_{ub}+\mu_u) a-\frac{\left(2 \mu_u^{3}-r_{ub}^{3}-r_{ub}^{2} \mu_u\right)}{\mu_u(r_{ub}-\mu_u)}\right) e^{-(r_{ub}+\mu_u) a} \right. \\
  &\quad \left. +\frac{r_{ub} \mu_u+2 \mu_u^{2}}{r_{ub}-\mu+u} e^{-\mu_u a}-\frac{r_{ub} \mu_u(r_{ub}+\mu)+r_{ub}^{3}}{\mu_u(r_{ub}-\mu_u)} e^{-r_{ub} a}\right) \nonumber \\
 &\times \left( \frac{\mu_u}{r_{ub}+\mu_u}\right)+	\left( \frac{\mu_u^{2}}{r_{ub}^{2}} e^{-(r_{ub}+\mu_u) a}(3 \mu_u+2 r_{ub}+r_{ub}(r_{ub}+\mu_u) a)\right.\\
&\quad \left. -\frac{\mu_u^{2}}{r_{ub}^{2}} e^{-\mu_u a}(3 \mu_u +2 r_{ub}-r_{ub}(r_{ub}+2 \mu_u) a)\right)
 \times \left( \frac{r_{ub}}{r_{ub}+\mu_u} \right). \nonumber
 \end{split}
 \vspace{-.9cm}
 \end{align}
}%
\normalsize
 Henceforth, to examine the worst-case scenario, i.e., the greatest \ac{AoI} peak, we consider the \emph{maximum severity of ruin} given the one-to-one relation between any M/G/1 system and a ruin probability.\footnote{The aggregation of our limited size queues and \ac{FCFS} computing queue constitutes an M/G/1 queue.} Thus, the ruin characterization of system reliability allows us to capture the effect of extreme events on the immersive real-time response. In other words, the maximum severity of ruin characterizes the severity of the worst peak \ac{QoPE} of the \ac{AR} user. Next, we derive the \ac{CDF} of the maximum severity of \ac{PAoI} ruin.
 \vspace{-.20cm}
 \begin{theorem}
The \ac{CDF} characterizing the maximum severity of \ac{PAoI}, for a \ac{PAoI} threshold of $z$ is given by:
\begin{align}
J_F(z)=Pr\left( M_F\leq z| a < \infty\right) =\frac{\Psi_F(a)-\Psi_F(a+z)}{\Psi_F(a)(1-\Psi_F(z))}, \label{eq1}\\
J_L(z)=Pr\left( M_L\leq z| a < \infty\right)=\frac{\Psi_L(a)-\Psi_L(a+z)}{\Psi_L(a)(1-\Psi_L(z))} \label{eq2}.
\end{align}
where $\Psi_F(a)$ is the \ac{CDF} of \ac{PAoI} over  M/M/1/2 \ac{FCFS} queues scheduling policies, $\Psi_L(a)$ is the \ac{CDF} of \ac{PAoI} over  M/M/1/2* \ac{LCFS} queues scheduling policies,  $M_F=\sup\left\lbrace A_F(t) | a< \infty \right\rbrace $,  $M_L=\sup\left\lbrace A_L(t) | a< \infty \right\rbrace $, $A_F(t)$ is the peak \ac{AoI} experienced at time $t$ over  M/M/1/2 \ac{FCFS} queues scheduling policies, and $A_L(t)$ is the peak \ac{AoI} experienced at time $t$ over  M/M/1/2* \ac{LCFS} queues scheduling policies. Thus, $M_F$ and $M_L$ are the worst-case peak \ac{AoI} for each case.
 \end{theorem}
\vspace{-.25cm}
\begin{IEEEproof}
See Appendix A	
\end{IEEEproof}
This result allows us to tractably assess the conditions during which worst-case \acp{PAoI} will not exceed a worst-case threshold $z$. The \acp{CDF} of both scheduling policies is a function of the \ac{THz} data rate and the queuing service rate. As long as the maximum severity of ruin \ac{PAoI} is maintained below $z$, a reliable, real-time response will be provided to the \ac{AR} user and the \ac{QoPE} will not be violated in extreme events. 
\vspace{-.35cm}
\subsection{Average Peak AoI}
In addition to the severity of ruin computed for the limited buffer size queues, we evaluate the average \ac{E2E} peak \ac{AoI} of our system.  Given that $Q_c$ is an M/M/1 queue, the average peak \ac{AoI} can be given by \cite{kosta2017age}: $\hat{A_{c}}=\frac{1}{\lambda_C}+\frac{1}{\mu_C}+\frac{\lambda_C(\mu_C+\mu_C^2)}{2(1-\rho_C)}$, where $\rho_C=\frac{\lambda_C}{\mu_C}$. Subsequently, the average \ac{E2E} peak \ac{AoI} can be given for each case as:
\begin{align}
	\hat{A}_{F}=&\frac{1}{\lambda_C}+\frac{1}{\mu_C}+\frac{\lambda_C(\mu_C+\mu_C^2)}{2(1-\rho_C)}\\ \nonumber
	&+ \sum_{u=1}^U\left( \frac{1}{r_{ub}}+\frac{3}{\mu_u}-\frac{2}{(r_{ub}+\mu_u)}\right), 
\end{align}
\vspace{-.25cm}
\begin{align}
\hat{A}_{L}&= \frac{1}{\lambda_C}+\frac{1}{\mu_C}+\frac{\lambda_C(\mu_C+\mu_C^2)}{2(1-\rho_C)}\\ \nonumber
&+ \sum_{u=1}^U\left( \frac{1}{r_{ub}}+\frac{1}{\mu_u}+\frac{r_{ub}}{(r_{ub}+\mu_u)^{2}}+\frac{r_{ub}}{\mu_u(r_{ub}+\mu_u)}\right) .
\end{align}
Here, in addition to our worst-case scenario analysis, the average \ac{PAoI} provides us with long term insights about \ac{PAoI} pertaining to a fair non extreme scenario. Moreover, here we also consider the computing queue in the aging process to provide us with the \ac{E2E} aging peak. 
	\begin{figure}[!t]
	\centering
	\includegraphics[scale=0.35]{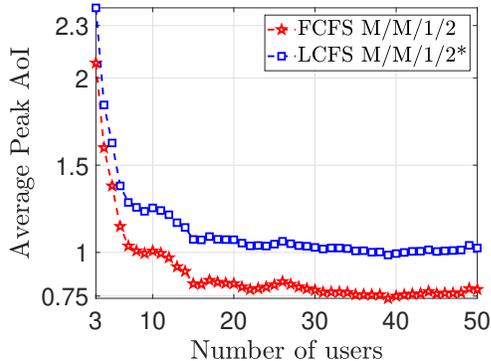}
	\caption{\small{Average peak AoI versus number of users}}
	\label{fig:averagepeakusers}
	\vspace{-.5cm}
\end{figure}
Next, in our simulation results, we compare the two suggested schemes and contrast the differences in their behavior from an expected and worst-case viewpoint.
\vspace{-.25cm}
\section{Simulation Results and Analysis}
\begin{figure}[!t]
	\centering
	\includegraphics[scale=0.35]{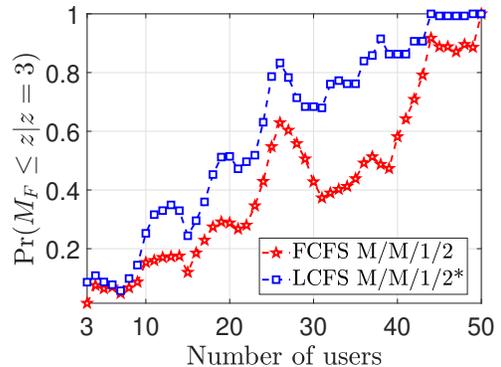}
	\vspace{-0.1cm}
	\caption{\small{Probability of worst-case peak AoI below $z=3$ versus number of users.}}
	\label{fig:WCSUSERS}
	\vspace{-.7cm}
\end{figure}

\begin{figure}[!t]
	\centering
	\includegraphics[scale=0.34]{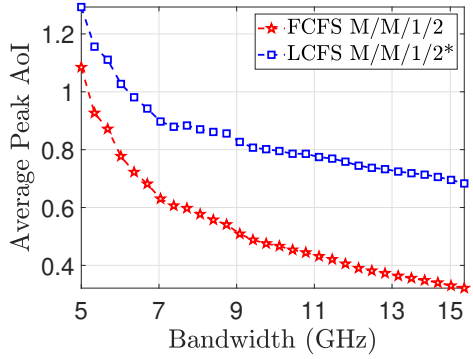}
	\caption{\small{Average peak AoI versus bandwidth}}
	\label{fig:averagepeakbw}
	\vspace{-.5cm}
\end{figure}

\vspace{-.15cm}
For our simulations, we consider the following parameters: $T_0= \SI{300}{K}$, $p=\SI{1}{W}$, $M= \SI{10}{Mbits}$, $f=\SI{1}{THz}$,  $W=\SI{10}{GHz}$, $K(f)= \SI{0.0016}{m^{-1}}$ with $1\%$ of water vapor molecules as in \cite{absorption} and we have chosen $\mu_i=\SI{5}{packets/s} $ $\mu_c=\SI{100}{packets/s}$ these values are chosen to comply with existing AR processing units such as the GEFORCE RTX 2080 Ti \cite{geforce}. The \acp{RIS} are deployed over the 4 walls of  an indoor area modeled as a square of size $\SI{50}{m}\times\SI{50}{m}$. \\
\indent Fig.~\ref{fig:averagepeakusers} and Fig.~\ref{fig:WCSUSERS} show the effect of the density of users on the freshness of information. We can see in Fig.~\ref{fig:averagepeakusers} that the peak average \ac{AoI} monotonically decreases with the number of users in the area. This is due to the fact that more packets are arriving simultaneously bringing fresher information into the network's server. In contrast to a typical \ac{E2E} delay analysis, an increase in the number of users would degrade the \ac{E2E} delay, while it improves the age as shown here. Moreover, the \ac{FCFS} M/M/1/2 queue's average \ac{PAoI} is $\SI{35}{\%}$ lower than the \ac{LCFS} M/M/1/2* queue for all densities. The gap between the two schemes increases as the number of users increases. \\ 
\indent Moreover, in Fig. \ref{fig:WCSUSERS}, we evaluate the distribution of worst-case peak \ac{AoI} below a threshold, where the threshold is set to $z=3$. This threshold was set based on the range of average \ac{PAoI} to contrast ruin behaviors to the expected one. Here, the number of users also positively impacts the performance, as the probability of the worst-case peak \ac{AoI} below $z$, is increasing with the number of users. Thus, as the density of users increases, the worst-case performance is further guaranteed to be below $z=3$. Nevertheless, even though the \ac{LCFS} M/M/1/2* queue had a higher average peak \ac{AoI}, it is shown here that it performs $\SI{12}{\%}$ better in terms of worst-case performance, and thus provides fresher information during \emph{extreme and dynamic events} when the number of users increases. In particular, \ac{LCFS} allow \ac{AR} users to deliver their more urgent requests while preempting their previous requests. This is especially useful during extreme events, that might delay urgent requests given that is incurring more latency to process the previous requests, and leading \ac{LCFS} M/M/1/2* queue will alleviate this issue.\\
\indent Fig.~\ref{fig:averagepeakbw} and Fig.~\ref{fig:WCSBW} show the prominent effect of the bandwidth on the freshness of information, for $U=15$ users. We can see in Fig. \ref{fig:averagepeakbw} that the peak average \ac{AoI} monotonically decreases with a higher bandwidth, this also corresponds to the fact that a higher number packets is arriving per unit time, thus ensuring information at the network's server. In here, the \ac{FCFS} M/M/1/2 queue's average peak \ac{AoI} is $\SI{45}{\%}$ lower than the \ac{LCFS} M/M/1/2* queue for all values of the bandwidth. The gap between the two schemes increases as the bandwidth increases.\\
\indent Moreover, in Fig.\ref{fig:WCSBW} we evaluate the distribution of worst-case peak \ac{AoI}, where the threshold is set to $z=3$. Here, and in contrast to the scenario of users density, the increase in bandwidth negatively impacts the ruin performance, i.e., the probability of a worst-case peak \ac{AoI} below $z$, is decreasing as the bandwidth increases. Clearly, a higher bandwidth changes the frequency range of \ac{THz}, thus, implying complications in terms of molecular absorption effect and range of operation, finally leading to extreme events with negative impact. Henceforth, increasing the bandwidth will increase the data rate and improve the performance of peak \ac{AoI} on average but will shift the worst-case peak \ac{AoI} to higher values. Here also, similarly to the density of the users, the \ac{LCFS} behaves better in terms of extreme events, but worse in terms of average peak \ac{AoI} performance.
\vspace{-.35cm}
\section{Conclusion}
\vspace{-.2cm}
In this paper, we have studied the \ac{AoI} of \ac{AR} services in a \ac{THz} cellular network employing \acp{RIS} as its \acp{BS}. We have performed a novel ruin-aware performance analysis that scrutinizes the \ac{PAoI} during extreme events. After deriving the \ac{PAoI} \acp{CDF}, we have used it to find the \ac{CDF} of the maximum severity of ruin for different scheduling policies. Moreover, we have also derived the average \ac{PAoI} for our considered model and compared the ruin performance analysis to the average \ac{PAoI} and contrasted the differences in the behavior. While introducing \ac{LCFS} preempting limited sized queues improved the ruin performance, \ac{FCFS} limited sized queues improved the average \ac{PAoI} performance.
\appendix
\vspace{-.25cm}
\subsection{Proof of Theorem 1}
\vspace{-.1cm}
\begin{IEEEproof}
	Given that the maximum severity of ruin is a function of the \ac{CDF} of ruin, we first need to integrate the results in \eqref{fcfs} and \eqref{preemption}. After some mathematical manipulations, we obtain the \acp{CDF} for the \ac{FCFS} M/M/1/2 queue and the \ac{LCFS} M/M/1/2* respectively:
	\begin{align}
	\vspace{-.35cm}
	\label{cdf_fcfs}
	&\varphi_F(a)=1-\frac{\mu_u^{3}}{(r_{ub}-\mu_u)^{2}(r_{ub}+\mu_u)} e^{-r_{ub} a}  \\
	\begin{split}
	&-\frac{r_{ub}}{2(r_{ub}-\mu_u)^{2}(r_{ub}+\mu_u)} e^{-\mu_u a}\times \left[\mu_u^{2} a^{2}(r_{ub}-\mu_u)^{2} \right.\\
	&\quad \left.+2 r_{ub} \mu_u(r_{ub}-\mu_u) a +2\left(r_{ub}^{2}-r_{ub} \mu_u-\mu_u^{2}\right) \nonumber \right],
	\end{split}
	\end{align}
	\vspace{-.25cm}
	\begin{align}
	\small
	\label{cdf_lcfs}
	\begin{split}
	&\varphi_L(a)= 1 -\frac{e^{-(r_{ub}+\mu_u) a}}{r_{ub}(r_{ub}+\mu_u)(r_{ub}-\mu_u)}\left(r_{ub}^{3}-3 \mu_u^{3} \right. \\
	&\quad \left.+r_{ub} \mu_u(r_{ub}+\mu_u)(1+\left( r_{ub}-\mu_u)\right) \right) +\frac{e^{-r_{ub} a}}{(r_{ub}+\mu_u)(r_{ub}-\mu_u)} \\
	&\times \left(r_{ub}^{2}+r_{ub} \mu_u+\mu_u^{2}\right)-\frac{e^{-\mu_u a}}{r_{ub}(r_{ub}+\mu_u)(r_{ub}-\mu_u)} \left(3 \mu_u^{3}+r_{ub} \right.\\
	&\quad \left. (r_{ub}-\mu_u)^2+r_{ub} \mu_u a\left(r_{ub}^{2}+r_{ub} \mu_u-2 \mu_u^{2}\right)\right). 
	\end{split}
	\vspace{-1cm}
	\end{align}
\vspace{-.005cm}
	Moreover, given that the \ac{AR} updates are \ac{i.i.d.}, the \ac{CDF} of the \ac{AoI} at the end of all limited buffer sized queues is given by:
	\begin{align}
	\vspace{-.15cm}
	\Psi_F(a)=\left( \varphi_F(a)\right)^u, \hspace{1cm} \Psi_L(a)=\left( \varphi_L(a)\right)^u.
	\end{align}
	Hence, the distribution of the maximum severity of ruin in each case can be given according to its definition in \cite{dickson2016insurance} by the result in \eqref{eq1} and \eqref{eq2} in Theorem 1.
\end{IEEEproof}	
\begin{figure}[!t]
	\centering
	\includegraphics[scale=0.34]{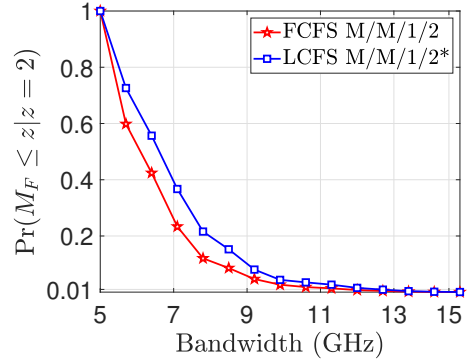}
	\caption{\small{Probability of worst-Case peak AoI below $z=2$ versus bandwidth}}
	\label{fig:WCSBW}
	\vspace{-.8cm}
\end{figure}
\bibliographystyle{IEEEtran}
\def\baselinestretch{0.83}
\bibliography{references}

\end{document}

%% file: acronyms.tex
\acro{D2D}{device-to-device}
\acro{SIR}{signal-to-interference-ratio}
\acro{SINR}{signal-to-interference-plus-noise-ratio}
\acro{PCP}{Poisson cluster process}
\acro{CoMP}{coordinated multi-point}
\acro{BS}{base station} 
\acro{MD-CoMP}{macrodiversity CoMP transmission}
\acro{MAC}{medium-access-control}
\acro{JT-CoMP}{joint transmission CoMP}
\acro{CoMP-JT}{coordinated multipoint joint transmission}
\acro{SBS}{small base station}
\acro{MDSD}{multiple devices to a single device}
\acro{MDS}{maximum distance separable}
\acro{SCN}{small cell network}
\acro{PPP}{Poisson point process}
\acro{TCP}{Thomas cluster process}
\acro{CSI}{channel state information}
\acro{PDF}{probability distribution function}
\acro{PMF}{probability mass function}
\acro{RV}{random variable}
\acro{i.i.d.}{independently and identically distributed}
\acro{MBMS}{multimedia broadcasting multicasting service}
\acro{EE}{energy efficiency}
\acro{HCP}{hard-core placement}
\acro{CCDF}{complementary cumulative distribution function}
\acro{CDF}{cumulative distribution function}
\acro{PC}{probabilistic caching}
\acro{RC}{random caching}
\acro{CPF}{caching popular files} 
\acro{PGFL}{probability generating functional}
\acro{KKT}{Karush-Kuhn-Tucker}
\acro{PGF}{point generating function}
\acro{SCA}{successive convex approximation}
\acro{HD}{high-definition}
\acro{FHD}{full-high-definition}
\acro{UHD}{ultra-high-definition}
\acro{VR}{virtual reality}
\acro{AR}{augmented reality}
\acro{5G}{fifth-generation}
\acro{QoS}{quality-of-service}
\acro{QoE}{quality-of-experience}
\acro{IoT}{Internet of Things}
\acro{MHCPP}{Matern hardcore point process}
\acro{LoS}{line-of-sight}
\acro{NLoS}{non-line-of-sight}
\acro{PSD}{power spectral density}
\acro{MEC}{mobile edge computing}
\acro{E2E}{end-to-end}
\acro{THz}{terahertz}
\acro{CLT}{central limit theorem}
\acro{HQ}{High Quality}
\acro{eMBB}{enhanced mobile broadband}
\acro{URLLC}{ultra reliable low latency communications}
\acro{mmWave}{millimeter wave}
\acro{EVT}{extreme value theory}
\acro{GEV}{generalized extreme value}
\acro{TVaR}{tail value at risk}
\acro{IoE}{Internet of Everything}
\acro{RL}{reinforcement learning}
\acro{GAN}{generative adversarial network}
\acro{HF}{high frequency}
\acro{RB}{resource block}
\acro{MIMO}{ multiple-input and multiple-output}
\acro{XR}{eXtended reality}
\acro{EM}{electromagnetic}
\acro{UE}{user equipment}
\acro{RIS}{reconfigurable intelligent surface}
\acro{IRS}{intelligent reflective surface}
\acro{HetNet}{heterogeneous network}
\acro{ML}{machine learning}
\acro{MLC}{machine learning for communications}
\acro{CML}{communications for machine learning}
\acro{RNN}{recurrent neural network}
\acro{MR}{mixed reality}
\acro{AoI}{age of information}
\acro{PAoI}{peak age of information}
\acro{RF}{radio frequency}
\acro{LCFS}{last come first served}
\acro{FCFS}{first come first served}
\acro{VaR}{value-at-risk}
\acro{CVaR}{conditional value-at-risk}
\acro{AVaR}{average value-at-risk}
\acro{ES}{expected shortfall}
\acro{QoPE}{quality of physical experience}
\acro{HRLLC}{high-rate and high-reliability low latency communications}